# Philosophical roots of the "eternal" questions in the XX-century theoretical physics

## V. Ihnatovych

Department of Philosophy, National Technical University of Ukraine "Kyiv Polytechnic Institute",
Kyiv, Ukraine
e-mail: V.Ihnatovych@kpi.ua

## Abstract

The evolution of theoretical physics in the XX century differs significantly from that in XVII-XIX centuries. While continuous progress is observed for theoretical physics in XVII-XIX centuries, modern physics contains many questions that have not been resolved despite many decades of discussion. Based upon the analysis of works by the founders of the XX-century physics, the conclusion is made that the roots of the "eternal" questions by the XX-century theoretical physics lie in the philosophy used by its founders. The conclusion is made about the need to use the ideas of philosophy that guided C. Huygens, I. Newton, W. Thomson (Lord Kelvin), J. K. Maxwell, and the other great physicists of the XVII-XIX centuries, in all areas of theoretical physics.

## 1. Classical Physics

The history of theoretical physics begins in 1687 with the work "Mathematical Principles of Natural Philosophy" by Isaac Newton. Even today, this work is an example of what a full and consistent outline of the physical theory should be. It contains everything necessary for such an outline – definition of basic concepts, the complete list of underlying laws, presentation of methods of theoretical research, rigorous proofs.

In the eighteenth century, such great physicists and mathematicians as Euler, D'Alembert, Lagrange, Laplace and others developed mechanics, hydrodynamics, acoustics and nebular mechanics on the basis of the ideas of Newton's "Principles". It is necessary to stress that initially, theoretical physics as a united logical system founded on the principles of mechanics covered only those phenomena that take place in the "weighty substance". For a long time since their emergence in the eighteenth century, theories of electricity, magnetism and heat existed separately from mechanics, as they presupposed some "weightless substance" – electric, magnetic, thermal – standing, as it was assumed, alongside the "weighted".

The discovery of the mechanical equivalent of heat (1842-1843) became a foundation for inclusion of thermal phenomena into mechanical physics. The mechanical (dynamic) theory of heat – assuming from the start that heat is mechanical movement of invisible body particles, atoms and molecules – began to develop. Electricity and magnetism were included into the general system of mechanical theoretical physics by J. C. Maxwell in his 1873 "Treatise on Electricity and Magnetism" based on the assumption that these phenomena represent mechanical motions of the ether.

In a speech delivered during general session at the opening of the First International Congress of Physics (1900), its chairman – president of the French Physical Society A. Cornu – said the following: "The mind of Descartes soars over modern physics, or rather, I should say, be is their luminary. The further we penetrate into the knowledge of natural phenomena, the clearer and the more developed becomes the bold Cartesian conception regarding the mechanism of the universe. There is nothing in the physical world but matter and movement" (cited from [1, p.10]).

Russian physicist B. B. Golitsyn wrote in 1893: "In general, physics undoubtedly tends towards mechanics now, trying to find a purely mechanical explanation of the various phenomena of Nature. The way it goes is quite clear, and even if it encounters significant difficulties on this way, – such as explaining the mechanical essence of electricity, – the success achieved in other areas leaves no doubt that it will prevail over all obstacles, and the hoped-for era will come when the various phenomena of the external physical world will be finally reduced to two basic principles, two fundamental principles of mechanics: matter and movement" [2, p.29-30].

Such confidence was shared by the overwhelming majority of physicists, and it was not accidental. Classical theoretical physics that reduced all physical phenomena to the mechanical motion of atoms and ether had explained almost all of the phenomena known in the late XIX century. On April 27, 1900, in a lecture entitled "Nineteenth Century Clouds over the Dynamical Theory of Heat and Light" [3], W. Thomson (Lord Kelvin) named only two unsolved problems in the theories of heat and light. «I. The first came into existence with the undulatory theory of light, and was dealt with by Fresnel and Dr. Thomas Young; it involved the question, How could the earth move through an elastic solid, such as essentially is the luminiferous ether? II. The second is the Maxwell–Boltzmann doctrine regarding the partition of energy" [3, p.1–2].

The development of theoretical Classical physics in XVII-XIX centuries can be compared to the construction of a building. Basic concepts and laws formulated by Newton served as foundation. This foundation was reaffirmed and expanded through the development of other formulations of Classical mechanics (Lagrangian mechanics, Hamiltonian mechanics). Based upon the foundation, the ground floor was built: the mechanics of a material point, mechanics of rigid body, etc.; theories of elasticity, plasticity, vibrations, the fluid dynamics, the theory of vortices, etc. On their basis, the next floor was built – the theory of sound, kinetic theory of gases, the theory of electromagnetism, as well as the theory of technical devices and machines – gyroscopes, ships, electromechanical transducers. All these theories form a single logical system and not only can be tested experimentally, but also serve as the basis for various practical calculations.

The creators of classical physics (C. Huygens, I. Newton, W. Thomson (Kelvin), J. Maxwell and others) were guided by the ideas of classical philosophy. The most important of them are: the statement of the existence of objective laws of nature, and the confidence in the possibility to discover them by means of thinking (dating back to the first philosophers and Aristotle); the rejection of groundless hypotheses, and the requirement to derive initial propositions of theories from the analysis of phenomena (from F. Bacon's philosophy); inadmissibility of attempts to explain the properties of objects by their inherent qualities instead of clarifying the mechanism of the phenomena, ideally –reduction of the properties to the movement of identical material points (from the philosophy by R. Decartes; dating back to Democritus and Epicurus).

## 2. Difficulties in the XX-century physics

The development of theoretical physics in the twentieth century is very different from that in XVII-XIX centuries. For decades, insuperable contradictions remain in the theory of relativity, quantum mechanics and theories created on their basis.

Some authors present this kind of statement: "Of course, nobody and nothing can refute, or even shake, the Theory of relativity and Quantum mechanics – these are the fundamentals of modern physics" [4, p. 340]. Others state: "There is no experimental evidence in support of the mathematically cumbersome Einstein's theory" [5, p.82]; "General Theory of Relativity is not only inconsistent logically from the physical point of view, it directly contradicts the experimental data on the equality of inertial and gravitational masses" [6, p.29].

Quantum mechanics is teeming with contradictions that are called paradoxes and discussed for decades (see, for example, [7, 8]).

Some of them receive the 2011 Nobel Prize in physics for the discovery of the accelerated expansion of the Universe [9], while others even deny the expansion of the Universe (e.g. [10, 11]).

As for the String theory, one can read: "It claims to be the one theory that unifies *all* the particles and *all* the forces in nature... Much effort has been put into string theory in the last twenty years, but we still do not know whether it is true" [12, p.XIV].

Similar examples can be cited for a long time.

If we compare XX century physics to a building, we have to say that not only its two bases are not linked into a single foundation, but each of them is full of contradictions (paradoxes). The number of "thunderstorm clouds" in the physical theory is ten times more today than in 1900. After decades of search and debates, even the direction of further research is not defined. Moreover, unlike the XVII-XIX century physics, it has no clear boundaries between the known and the unknown, the reliable and doubtful, true and false.

Today, more and more writings are dedicated to the crisis of physics, the crisis of physical-mathematical community, the "end of science" (e.g. [12-14]).

Since nothing like this has happened in physics prior to the twentieth century, it is possible to conclude that the crisis is caused by something that is common to the theories created in the twentieth century, – and how they differ from the theories of Classical physics. And this "something" is philosophy that has been put in their foundation.

## 3. Physics of the XX century denies objective reality and objective laws

It is known that the creators of Theory of relativity and Quantum mechanics were guided by the ideas of positivist philosophy, Ernst Mach's philosophy in particular.

Einstein wrote: "Theory of relativity arose from attempts to improve, on the basis of the economy of thought, the foundations of physics that existed in the beginning of the century" [15]; "I was helped, either directly or indirectly, in particular by Hume and Mach's works" [15].

Einstein provides his definition of scientific aims, borrowed from Mach's philosophy, at the very beginning of his book "The Meaning of Relativity": "The object of all science, whether natural science or psychology, is to co-ordinate our experiences and to bring them into a logical system" [17, p.1].

This definition is imported from Mach's philosophy. Mach wrote that the goal of physics is "to determine the laws of connection between sensations (perceptions)..." ["Die Gesetze des Zusammethanges der Empfindungen (Wahrnehmungen) aufzufinden"] [18, p.58]. Mach denied the existence of reality which is not available to the senses.

Special theory of relativity is based upon the denial of objectivity of space and time: distances and time intervals can only be determined through certain procedures. Einstein introduced the measurement of lengths and time intervals by using light signals as such procedures, and postulated that the speed of light in all reference frames is the same. The result was the dependence of length and time intervals on the speed of observers.

The ideas of positivist philosophy constitute the basis of Quantum mechanics.

"…The usual interpretation of QM (Quantum mechanics), as found e.g. in the classical treatises of von Neumann (1932) and Dirac (1958), as well as in standard textbooks such as those of Bohm (1951) and Landau and Lifshitz (1958), has been cast in the spirit and letter of the early logical positivism…" [19, p.87].

By that time, Einstein's philosophical views changed, and he started to criticize the assumptions of Quantum mechanics.

A. Einstein wrote in 1938: "...Nowadays, the subjective and positivist view prevails. Proponents of this view claim that the consideration of nature as an objective reality is an outdated prejudice. That's what theorists involved in Quantum mechanics put to their merit" [20].

In his article "Introductory remarks about the basic concepts" (1953), Einstein wrote:

"I never doubt the fact that modern Quantum theory (or, rather, "Quantum mechanics") stands in the most comprehensive agreement with experiment, as long as "material point" and "potential energy" are the elementary concepts that form the description basis. However, I find unsatisfactory the interpretation that is given to "ψ-function". In any case, my understanding of the situation is based on the following statement that is emphatically rejected by major contemporary theorists:

*There is something like a "real state"* of a physical system that exists objectively, independently of any observation or measurement – which can be described in principle by using the available tools of physics [which adequate tools should be used for this, and therefore, what fundamental concepts should be used, in my opinion, is still unknown. (Material point? Field? Any other way of description, which is yet to be found?)] This thesis about the reality has no clear meaning in itself, given its "metaphysical" nature – it is only of a *programmatic* character. Everyone, however, – including the theorists involved in Quantum mechanics, – adheres to this provision about the reality until the basics of Quantum mechanics are not discussed. Nobody, for example, has any doubt that the center of gravity of the Moon, at some preassigned time, occupies a definite position even if there is no (real or potential) observer. If one discards this arbitrary thesis about the reality, considered in purely logical terms, it will be very difficult to avoid solipsism" [21].

Einstein once illustrated the latter conclusion with the question: "...moon exists only if I look at it"? [22, p.5].

This was the reason why the famous dispute between A. Einstein and N. Bohr on the foundations of Quantum mechanics lasted for so long and resulted in nothing: Einstein represented Classical philosophy and physics in this argument, while Bohr represented the positivist philosophy. In the dispute about fundamentals (interpretation) of Quantum mechanics, Einstein believed that the theory should get closer to the reality standing behind the facts, while Bohr denied the existence of such reality.

Any course on Quantum mechanics includes arguments justifying the "inapplicability of laws and concepts of Classical physics to the microworld" and the "need to abandon classical prejudices". In reality, these arguments represent nothing more than hypnosis sessions to convince the reader to abandon the idea of objectiveness of the research subject and the laws of nature – the idea that was central not only to Classical physics, but to the whole of the rationalist European philosophy and science. To agree with the starting point of Quantum mechanics, one has to reject the distinction between something that exists in reality and something that exists in the imagination, the distinction inherent to any sensible individual. Therefore, quantum theorists strongly disparage common sense but they do not know how to answer Einstein's question about the Moon – they are still debating it in scientific journals (see, for example, [23]).

## 4. Physics of the XX century has revived the scholastics

A major guiding idea for Classical physics was the Decartes' idea about the necessity to reduce all the phenomena to movement of homogenous matter, directed against the scholastic explanations of the properties of objects by their special quality or nature (e.g. heavy bodies fall because they are supposed to move down). It was on the basis of this idea that the molecular-kinetic theory of gases was created and developed consistently. The most simple mechanical model of gas (molecules of gas are represented by moving material points) enabled the derivation of the empirically known Boyle-Mariotte law, and also to calculate the velocity of molecules. The latter turned out to be hundreds of meters per second, which contradicted the experimental data on the diffusion of gases. The contradiction was solved when the size and collision of molecules were taken into account. The discrepancy between the theoretical and empirically determined heat capacity was eliminated by taking into account rotation and vibration of the molecules. Assuming that the different properties of gases were due to movements of the same molecules, quantitative relationships were set up in the kinetic theory of gases, for example, between the coefficients of thermal conductivity, diffusion and viscosity of the gas.

It seems clear that the theoretical link between the above mentioned factors would be impossible to find without the mechanical model of gas. The same kind of impossibility is shown by the unsuccessful 30-year-long effort by Einstein working on the unified field theory: he tried to create a unified theory of electromagnetic and gravitational interactions while abandoning the assumption of these fields being manifestations of a certain environment.

The rejection of Carthesianism led naturally to a revival of the kind of scholasticism in physics that the founders of new European philosophy and classical physics were fighting against. Today, theories of elementary particles are dealing with "strange", "charm", "color" and other "inborn qualities". By the way, the standard method of modern theoretical physics – derivation of consequences from the postulates of Quantum mechanics and Theory of relativity – comes from the arsenal of medieval scholastics which derived all of the conclusion about the world from the Bible and the works by Aristotle, without resorting to experimentation.

In recent years, the growing number of authors are writing about the crisis of theoretical physics, but usually they can not specify the way out, because they do not even mention the materialistic and idealistic trends in physics and repeat uncritically the idealists' allegations of limitations of Classical physics without having any clue about the methods by which Classical physics was created. Thus, they try to overcome the crisis in the theory without using the ideas by which the Western philosophy and science was developing for twenty-five centuries – ideas that guided Newton, J. K. Maxwell, W. Thomson, Helmholtz, Hertz, J. J. Thomson and other classicists. Nobody even sets up the question: why elementary particles should certainly be described as fluctuations of zero-dimensional segments in 10 (or more) dimensional space, but not as gas-like vortices in the usual three-dimensional ether?

**Conclusions**

As long as the idea of denying objective reality, borrowed from the positivist philosophy, lies in the foundations of Theory of relativity and Quantum mechanics, it is impossible to overcome the crisis of physics by some partial improvements in these theories. This is convincingly demonstrated by the history of physics of the twentieth century. It will only be possible to find a way out of the crisis if the physical theory is based on the principles of classical physics. The most important of those is the proposition about the real (outside of consciousness - in space and time) existence of objects studied by physics, the reflection of inherent patterns in the theories, and explanation of various phenomena by the mechanical motion of mediums and particles.